\documentclass[aps,pra,twocolumn,groupedaddress]{revtex4}
\usepackage{graphicx}
\usepackage{amsmath}
\usepackage{amsfonts,amssymb}

\newcommand{\rb}[1]{\left( #1 \right)}

\newcommand{\beq}{\begin{eqnarray}}
\newcommand{\eeq}{\end{eqnarray}}

\newcommand{\ew}[1]{\langle #1 \rangle}
\begin{document}


\title{Phase Transitions in Generalised Spin-Boson (Dicke) Models}

\author{Clive Emary}
\email{emary@lorentz.leidenuniv.nl}
\affiliation{ Instituut--Lorentz, 
              Universiteit Leiden,
              P. O. Box 9506 RA Leiden,
              The Netherlands}
\author{Tobias Brandes}
\email{brandes@dirac.phy.umist.ac.uk}
\affiliation{ Department of Physics,
           UMIST,
           P.O. Box 88,
           Manchester 
           M60 1QD, 
          U. K.}


\date{\today}

\begin{abstract}
We consider a class of generalised single mode Dicke Hamiltonians with
arbitrary boson coupling in the pseudo-spin $x$-$z$ plane. 
We find exact solutions in the thermodynamic, large-spin  limit 
as a function of the coupling 
angle, which allows us to continuously move
between the
simple dephasing and the original Dicke Hamiltonians. 
Only in the latter case
(orthogonal static and fluctuating couplings), does the parity-symmetry 
induced quantum phase transition occur.
\end{abstract}


\maketitle
\section{Introduction}
Spin-Boson models appear in many areas of  physics and 
are essential ingredients in theoretical quantum optics \cite{Walls}
(light-matter interaction), nuclear physics \cite{KM91}, 
quantum chaos \cite{Haake},  and quantum dissipation \cite{Weiss}. 
The spin algebra can be used to describe single ($j=N/2=1/2$) or
many $N>1$ two-level systems where, in the simplest case, the interaction
is with but a single bosonic mode ($a, a^{\dagger}$). Specific examples 
include
cavity quantum electrodynamics and, more recently, `phonon cavity
quantum  dynamics' of electrons interacting with
single phonon (oscillation) modes in nano-electromechanical systems 
\cite{Paretal00,CR98,Blietal00,BL03} such as
freestanding quantum dots or  `molecular transistors'.

A common feature of spin-boson models is that in general they are
non-integrable, with 
exact solutions available only for very specific cases. Examples of 
the latter are
simplified `dephasing models', where the spin couples to both the boson and 
static field via only one of its components (usually chosen as $J_z$).
Another example where  exact solutions can be obtained 
is in the large spin limit $j\to \infty$ where
bosonic representations of spin Lie algebras \cite{KM91} 
have been known for a long time; an early example being the 
Holstein-Primakoff transformation \cite{HP49}.

In this paper, we further explore the large-spin limit by
starting from the most general, single- mode, spin boson Hamiltonian with
linear coupling of all ($x$, $y$, $z$) spin components to a static {\em and} 
a fluctuating (bosonic) term. For the 
specific case of the coupling of orthogonal ($x$ and $z$) spin components
to the static and the fluctuating term (Dicke model), we have previously 
found \cite{EB03,EB03a,LEB03}
intriguing  connections between quantum chaos, entanglement, and
the emergence of an instablity-induced quantum phase transition in the 
limit of large spins $j\to \infty$. 
Here, our main result will be that, surprisingly, this instability
and the related parity-symmetry breaking of the ground-state wave functions only
appears for `orthogonal' coupling. The Dicke Hamiltonian \cite{Dic54} (Rabi-Hamiltonian for spin $1/2$) 
and its canonical equivalents  therefore seem to be in a  `distinguished' class of Hamiltonians with very
pronounced properties. It should be mentioned from the very beginning, however, that
this distinction is most visible in the strong coupling regime.

\section{The model and its solution\label{solution}}
We start from a generic model  Hamiltonian 
\begin{eqnarray}\label{Hamiltonian}
  H = \omega a^{\dagger} a + \left ({\bf \Omega } + a^{\dagger} {\bf \Lambda }
+ a {\bf \Lambda }^{\dagger} \right) {\bf J},
\end{eqnarray}
describing the simplest coupling between 
Heisenberg-Weyl ($1,a,a^{\dagger}$) and the spin algebras 
$J_x=\frac{1}{2}(J_++J_-)$,
$J_y=\frac{1}{2}(J_+-J_-)$, $J_z$,  with 
\begin{eqnarray}
  [J_z,J_{\pm}]=\pm J_{\pm},\quad [J_+,J_-]=2J_z.
\end{eqnarray}
In Eq.(\ref{Hamiltonian}), ${\bf \Omega }$ is a real and ${\bf \Lambda }$
a complex three-dimensional vector. Special cases of  Eq.(\ref{Hamiltonian}) are
the Rabi or the Dicke Hamiltonian \cite{Allen} (${\bf \Omega }=\Omega{\bf e}_z$,
${\bf \Lambda }={\bf \Lambda }^{\dagger}=\Lambda {\bf e}_x$),
the simple dephasing Hamiltonian \cite{PSE96,RQJ02,YE03}
(${\bf \Omega }=\Omega{\bf e}_i$,
${\bf \Lambda }={\bf \Lambda }^{\dagger}=\Lambda {\bf e}_i$ with $i=x,y$, or $z$),
the Jaynes-Cummings Hamiltonian \cite{Walls}
(${\bf \Omega }=\Omega{\bf e}_z$,
${\bf \Lambda } = \Lambda[{\bf e}_x - i {\bf e}_y]$), and
the one-mode version of the dissipative spin-boson (tunneling electron) 
Hamiltonian
\cite{Legetal87,Weiss,Brandesddot,BL03}
(${\bf \Omega }=\omega_0 {\bf e}_z + T {\bf e}_x$,
${\bf \Lambda }={\bf \Lambda }^{\dagger}=\Lambda {\bf e}_z$), where we denoted
the unit vectors as ${\bf e}_i$, $i=x,y,z$.
The $j=1/2$ variant of  Eq.~(\ref{Hamiltonian}) with ${\bf \Omega }=\omega_0 {\bf e}_z + T {\bf e}_x$
and ${\bf \Lambda }= a {\bf e}_x + i b  {\bf e}_y$ appears
in quasi one-dimensional quantum wires in the $x$-$y$ plane
in a constant magnetic field $B{\bf e}_z$ for an electron gas with 
spin-orbit interactions (Rashba Hamiltonian) \cite{DB03}.

In the following, we restrict ourselves to
$
  {\bf \Lambda }={\bf \Lambda }^{\dagger}
$
and therefore consider the Hamiltonian
\begin{eqnarray}\label{Hamfinal}
  H =  \omega a^\dag a + \mathbf{\Omega} \cdot \mathbf{J}
  + \rb{a^\dag + a} \mathbf{\Lambda} \cdot \mathbf{J}
\end{eqnarray}
parameterised by  {\em two} real three-dimensional vectors given by
\begin{eqnarray}
  \mathbf{\Omega} = \rb{\Omega_x,\Omega_y,\Omega_z}
  ;~~~
  \mathbf{\Lambda} =
  \frac{2}{\sqrt{2 j}}\rb{\lambda_x,\lambda_y,\lambda_z}
\end{eqnarray}
where $1/\sqrt{2j}$ is inserted to ensure correct
scaling in the thermodynamic limit, and the factor of 2 is for 
later convenience.
This Hamiltonian is invariant 
under a rotation about the $z$-axis, under which $J_z\rightarrow J_z$
and $J_x\rightarrow - J_x$, consequently, we shall only discuss the
parameter range $0\le \theta\le \pi$. Note that the more general case, 
Eq.(\ref{Hamiltonian}),
in general would leave three real, linearly independent three-dimensional 
vectors. 
The analysis would then be very similar to the following, though more 
cumbersome, which is why we restrict ourselves
to the model Eq.(\ref{Hamfinal}).

We begin by rotating our co-ordinate axes so that we work in the $x$-$y$
plane, with the coupling-vector $\mathbf{\Lambda}$ aligned along the 
$x$-axis.  This gives us the form with which we shall work:
\begin{eqnarray}
  H = \omega a^\dag a + \Omega \rb{ J_x \cos \theta + J_z \sin \theta}
  + \frac{2 \lambda}{\sqrt{2j}}\rb{a^\dag + a} J_x
 \label{oham}
\end{eqnarray}
In deriving exact solutions for this model in the thermodynamic limit,
we shall follow the general procedure introduced for the Dicke model 
in Ref. \cite{EB03a}.

First we employ the Holstein--Primakoff representation of the
angular momentum operators \cite{HP49},
$J_+ = b^\dagger \sqrt{2j - b^\dagger b}$,
$J_- = \sqrt{2j - b^\dagger b}~ b$,
$J_z = \rb{b^\dagger b - j}$.  With
$J_x = \frac{1}{2} \rb{J_+ + J_-}$, 
substitution gives us
\begin{eqnarray}
  H &=& \omega a^\dag a 
  + \frac{\Omega}{2}\cos\theta
    \rb{b^\dag \sqrt{2j - b^\dag b} + \sqrt{2j - b^\dag b}~ b} \nonumber\\
 & + &\Omega\sin\theta \rb{b^\dag b -  j} \\
  &+& \frac{\lambda}{\sqrt{2j}}\rb{a^\dag + a}
    \rb{b^\dag\sqrt{2j - b^\dag b} + \sqrt{2j - b^\dag b}~ b}\nonumber
\end{eqnarray}
We next displace the oscillator modes $a \rightarrow a + \sqrt{\alpha}$ and 
$b \rightarrow b - \sqrt{\beta}$, where $\alpha$ and $\beta$ are assumed to
be
of the order of $j$.  This leads to
\begin{widetext}
\begin{eqnarray}
  H &=& \omega \rb{a^\dag a + \sqrt{\alpha}\rb{a^\dag + a} + \alpha}
  + \frac{1}{2}\Omega\cos\theta \sqrt{k}
    \rb{b^\dag \sqrt{\eta} + \sqrt{\eta}~ b - 2 \sqrt{\beta}\sqrt{\eta}}
  \nonumber \\
  &+& \Omega\sin\theta \rb{b^\dag b -\sqrt{\beta} \rb{b^\dag + b}+\beta -  j}
  + \lambda \sqrt{\frac{k}{2j}}\rb{a^\dag + a + 2 \sqrt{\alpha}}
    \rb{b^\dag \sqrt{\eta} + \sqrt{\eta}~ b - 2 \sqrt{\beta}\sqrt{\eta}}, 
\end{eqnarray}
\end{widetext}
where
\begin{eqnarray}
  k = 2j-\beta;~~~
  \eta = 1 - \frac{b^\dag b - \sqrt{\beta} \rb{b^\dag + b}}{k}
\end{eqnarray}
We now proceed to the thermodynamic limit, by taking 
$j\rightarrow \infty$ and neglecting terms with powers of $j$ in the 
denominator.  This yields
\begin{widetext}
\begin{eqnarray}
  H^{j\rightarrow \infty} &=& 
  \omega a^\dag a
  + \rb{
        \Omega \sin \theta 
        + 2\lambda\sqrt{\frac{\alpha\beta}{2jk}} 
        + \frac{\Omega \cos \theta}{2}\sqrt{\frac{\beta}{k}}
       } b^\dag b
  + \rb{\omega\sqrt{\alpha}-2 \lambda \sqrt{\frac{\beta k}{2 j}}} 
      \rb{a^\dag + a}
  \nonumber \\
  &&+ \rb{
          4\lambda\sqrt{\frac{\alpha }{2 j k}}\rb{j-\beta}
          - \Omega \sin \theta \sqrt{\beta}
          + \Omega \cos \theta \rb{\frac{j-\beta}{\sqrt{k}}}
         } 
      \rb{b^\dag + b}
  \nonumber \\
  &&+ \rb{
          \frac{\lambda}{2k}\sqrt{\frac{\alpha \beta}{2 j k}} \rb{2k+\beta}
          +\frac{1}{4}\Omega \cos\theta \sqrt{\frac{\beta}{k}}
             \rb{1+\frac{\beta}{2 k}}
         } \rb{b^\dag + b}^2
  + 2\lambda\sqrt{\frac{1}{2jk}}\rb{j-\beta} \rb{a^\dag + a} \rb{b^\dag + b}
  \nonumber \\
  &&+ \Omega\sin\theta\rb{\beta-j} + \omega \alpha 
    - \Omega\cos\theta\sqrt{k\beta}
    -\lambda \sqrt{\frac{\alpha \beta}{2jk}}\rb{1+4k}
    -\frac{1}{4}\Omega\cos\theta\sqrt{\frac{\beta}{k}}.
  \label{fullTDLHam}
\end{eqnarray}
\end{widetext}
The two terms linear in bosonic operators 
can be eliminated by choosing the parameters $\alpha$ and $\beta$ 
such that 
\begin{eqnarray}
\sqrt{\alpha} = \frac{2 \lambda}{\omega}\sqrt{\frac{k\beta}{2j}},
\end{eqnarray}
and $\beta$ is determined by
\begin{eqnarray}
\lefteqn{   \frac{4\lambda}{k}\sqrt{\frac{\alpha k}{2 j}}\rb{j-\beta} 
          - \Omega \sin \theta \sqrt{\beta}  }\nonumber\\
       &&   + \frac{1}{2}\Omega \cos \theta \sqrt{k}\rb{1-\frac{\beta}{k}}  
  =0.
\end{eqnarray}
Substituting the value of $\alpha$ into this equation and simplifying, 
we obtain the following equation for $\sqrt{\beta}$,
\begin{eqnarray}
  \frac{4 \lambda^2}{\omega} \frac{j-\beta}{j}\sqrt{\beta}
  -\Omega\sin\theta \sqrt{\beta}
  +\Omega\cos\theta \frac{j-\beta}{\sqrt{2j -\beta}}
  =0.
  \label{beta}
\end{eqnarray}
This equation is exactly soluble for $\sqrt{\beta}$, but the resulting 
form is extremely unwieldy.
The solutions of this equation
for arbitrary parameters are cumbersome, and will not be reproduced here.  
In a few specific cases, to be elucidated later, compact expressions can 
be found.
With the elimination of the linear terms, our Hamiltonian assumes the form
\begin{eqnarray}
  H &=& \omega a^\dag a 
    + \tilde{\omega} b^\dag b
    + s \rb{b^\dag + b}^2 \nonumber\\
    &+&  r \rb{a^\dag + a}\rb{b^\dag + b}
    + j E_G + k',
  \label{Hpar}
\end{eqnarray}
where the constants may be inferred by comparison with 
Eq. (\ref{fullTDLHam}), with appropriate values 
of $\alpha$ and $\beta$.
Hamiltonians of this form are analytically soluble via a unitary 
transformation, and since an example of this process was given in 
\cite{EB03a}, we shall not go into the details here.  Suffice to say that 
after a Bogoliubov transformation of the bosonic operators, the Hamiltonian
becomes diagonalised
\begin{eqnarray}
H = \varepsilon_+ c_+^\dag c_+ + \varepsilon_- c_-^\dag c_- + j E_G + k,
\end{eqnarray}
where we have introduced the excitation energies of the system,
$\varepsilon_\pm$, and where $E_G$ is the scaled ground-state energy 
(scaled with $j$) 
and $k$ is an unimportant constant of the order unity.  In 
terms of the parameters introduced in Eq. (\ref{Hpar}), 
the excitation energies are given by
\begin{eqnarray}
\lefteqn{  \varepsilon_\pm^2 = } \\
&& \frac{1}{2}
    \rb{
        \omega^2 + \tilde{\omega}^2 + 4 \tilde{\omega} s 
        \pm \sqrt{ 
                  \rb{\tilde{\omega}^2 + 4 \tilde{\omega} s -\omega^2}^2 
                  + 16 r^2 \omega \tilde{\omega}
                 }
       }. \nonumber
\end{eqnarray}
and, in terms of $\beta$, the ground-state energy is given by
\begin{eqnarray}
  j E_G &=& \Omega \sin \theta \rb{\beta - j}\\
      & + &\frac{2 \lambda^2}{j\omega} \beta \rb{2j-\beta}
       -\Omega \cos \theta \sqrt{\beta\rb{2j-\beta}}. \nonumber
\end{eqnarray}
The general scheme in which we proceed from here is to solve Eq. (\ref{beta})
for $\beta$, and then use this value to compute the excitation 
and ground-state energies.  Before considering the problem with 
arbitrary parameters however, we will focus on two special cases, which 
will explain many of the features of the general solution.
It should be pointed out that not all solutions of Eq. (\ref{beta}) are 
physically valid, and by considering the following cases we 
shall determine the criteria for selecting
valid solutions.

\section{Specific Limits}

\subsection{The Dicke model: $\theta = \pi/2$}
In the case where the interaction and spin vectors are perpendicular we 
obtain the Dicke model:
\begin{eqnarray}
  H_{\pi/2} = \omega a^\dag a + \Omega J_z
  + \frac{2 \lambda}{\sqrt{2j}}\rb{a^\dag + a} J_x.
\end{eqnarray}
In this limit there exists a conserved parity $\Pi$
such that $\left[H,\Pi\right]=0$, given by
\begin{equation}
  \Pi= \exp\left\{i\pi \hat{N}\right\};~~~
  \hat{N} = a^\dagger a + J_z+j,
  \label{parity}
\end{equation}
where $\hat{N}$ is the ``excitation number'' 
and counts the total number of excitation quanta in the system.
$\Pi$ possesses two eigenvalues, $\pm 1$, depending on
whether the number of quanta is even or odd.  

For the Dicke Hamiltonian, the equation for determining $\sqrt{\beta}$
becomes
\beq
 \sqrt{\beta}
    \rb{ 4 \lambda^2 \rb{j-\beta} -j\Omega~\omega}
  =0.
\label{betaDH}
\eeq
The simplest solution sets $\sqrt{\beta}=\sqrt{\alpha}=0$, which
gives rise to the effective Hamiltonian
\begin{equation}
  H^{(1)}_{\pi/2} = \omega_0 b^\dagger b + \omega a^\dagger a
   + \lambda \rb{a^\dagger + a}\rb{b^\dagger + b} - j \omega_0,
  \label{lcDHjinf}
\end{equation}
which has the excitation energies
\begin{equation}
  {\varepsilon^{(1)}_{\pm}}^2 = \frac{1}{2}\left\{ \omega^2 + \omega_0^2 
  \pm \sqrt{\rb{\omega_0^2 - \omega^2}^2 
  + 16 \lambda^2 \omega \omega_0}\right\}\label{lcepspm}.
\end{equation}
and the ground-state energy $E^{(1)}_G  = -j\omega_0 $.  The excitation 
energy $\varepsilon^{(1)}_-$ remains real provided that
$\lambda \le \lambda_c = \sqrt{\omega \omega_0}/2$, and this 
demarcates the range of validity of this solution.  The appearance 
of an imaginary part
of an eigenenergy is one of our criteria for distinguishing
between valid and invalid solutions of Eq. (\ref{beta}).

The remaining two solutions of Eq. (\ref{betaDH}) are given by the 
displacements
\begin{eqnarray}
  \sqrt{\alpha} = \pm \frac{2 \lambda}{\omega}\sqrt{\frac{j}{2}\rb{1 - \mu^2}},
\quad 
  \sqrt{\beta} = \pm\sqrt{j \rb{1- \mu}},
  \label{abdet}
\end{eqnarray}
where we have defined $\mu \equiv \frac{\omega \omega_0}{4 \lambda^2}$ 
$= \frac{\lambda_c^2}{\lambda^2}$.  The Hamiltonians
obtained with these solutions (one for each sign) are identical and have the 
same excitation energies
\begin{eqnarray}
{\varepsilon_\pm^{(2)}}^2 = \frac{1}{2} \left\{
  \frac{\omega_0^2}{\mu^2} +  \omega^2
  \pm \sqrt{\left[ \frac{\omega_0^2}{\mu^2} - \omega^2 \right]^2 
  + 4\omega^2 \omega_0^2} \right\},
\end{eqnarray}
and ground-state energy,
\beq
  E^{(2)}_G = -\left\{
  \frac{2 \lambda^2}{\omega}+ \frac{\omega_0^2 \omega}{8 \lambda^2}  
  \right\},
  \label{gse}
\eeq
and we thus see that these two solutions are completely 
degenerate.  By considering the reality of $\varepsilon_-^{(2)}$, we 
conclude that these second two solutions are only valid providing 
$\lambda \ge \lambda_c$.

As described in \cite{EB03} and to be discussed later, 
the existence of these different solutions,
one with zero displacement, and two with finite and opposite displacements, 
describes a quantum phase transition in the Dicke model, which occurs 
at the critical coupling $\lambda_c$.  The nature of this QPT is such 
that the parity symmetry becomes broken above $\lambda_c$, which 
explains the appearance of the two degenerate, broken symmetry 
solutions.
\subsection{One-dimension: $\theta=0$}
With interaction and spin aligned, the full Hamiltonian
of Eq. (\ref{oham}) becomes
\begin{eqnarray}
  H_0 \rb{j} = \omega a^\dag a + \Omega J_x
  + \frac{2 \lambda}{\sqrt{2j}}\rb{a^\dag + a} J_x.
  \label{H0}
\end{eqnarray}
This Hamiltonian is integrable for arbitrary $j$ since
its eigenstates are clearly also eigenstates of $J_x$,
which allows us to replace the operator with its eigenvalue 
$m_x = -j,-j+1\ldots, j-1,j$, such that
\begin{eqnarray}
  H_0\rb{j} = \omega a^\dag a + \Omega m_x 
  + \frac{2 \lambda}{\sqrt{2j}}\rb{a^\dag + a} m_x.
\end{eqnarray}
This leaves us with a single-mode bosonic Hamiltonian 
which may be diagonalised 
via a simple displacement 
$a \rightarrow a - \rb{2 \lambda m_x} / \rb{\sqrt{2j}\omega}$.  This results 
in the diagonal form
\begin{eqnarray}
  H_0\rb{j} 
  = \omega a^\dag a + \Omega m_x - 2 \frac{\lambda^2 m_x^2}{j \omega},
\end{eqnarray}
which has the energy 
\begin{eqnarray}
  E_{n,m_x}= \omega n + \Omega m_x - 2 \frac{\lambda^2 m_x^2}{j \omega}.
\end{eqnarray}
We proceed to the thermodynamic limit by writing $m_x = k_x - j$, 
and neglecting terms with $j$ in the denominator.  Whence,
\begin{eqnarray}
 E_{n,k_x}^{j\rightarrow \infty}=
  \omega n 
  + \rb{\Omega + 4 \frac{\lambda^2}{\omega}} k_x
  - j \rb{\Omega + 2\frac{\lambda^2}{\omega}},
\end{eqnarray}
from which we immediately see that the excitation energies are  
$\varepsilon_- = \omega$  and 
$\varepsilon_+ = \Omega + 4 \lambda^2 / \omega $, 
and the scaled ground-state energy is 
$E_G = - \rb{\Omega + 2\lambda^2/\omega}$.  

We now seek to obtain this results using the general procedure 
outlined in section \ref{solution}.
The equation for the determination of $\beta$ becomes
\beq
  \rb{j-\beta} 
  \rb{j \Omega \omega + 4 \lambda^2 \sqrt{\beta \rb{2j -\beta}}}
  =0.
  \label{betadet0}
\eeq
Setting the second factor in this expression to zero leads
to values of $\sqrt{\beta}$ and $\sqrt{\alpha}$ which give rise to 
complex excitation 
energies for all parameter values.  
These solutions are unphysical  and we discard them as we
did for the Dicke model.
Considering the other solution, we have $\beta =j$, 
which gives $\sqrt{\beta} = \pm \sqrt{j}$ and
$\sqrt{\alpha} = \pm \frac{\lambda}{\omega} \sqrt{2j}$.  With these choices,
the Hamiltonian of Eq. (\ref{Hpar}) becomes
\beq
 H_0^{j \rightarrow \infty} &=& 
 \omega a^\dag a \nonumber\\
  &+& \rb{
      \frac{2\lambda^2}{\omega} \pm \frac{\Omega}{2}
    }
    \rb{
      b^\dag b + \frac{3}{4} \rb{b^\dag + b}^2 - \frac{1}{2}
    }  \nonumber\\
 & - &j 
    \rb{
      \frac{2 \lambda^2}{\omega} \pm \Omega
    }.
\eeq
Note that the two modes are now decoupled.  The $b$-mode may be 
diagonalised via the squeezing transformation, 
\begin{eqnarray}
  b \rightarrow \frac{1}{\sqrt{1-\sigma^2}} \rb{b^\dag + \sigma b},\quad 
b^\dag \rightarrow \frac{1}{\sqrt{1-\sigma^2}} \rb{b + \sigma b^\dag},
\end{eqnarray}
with the squeezing parameter $\sigma = -1/3$.  In this way we arrive at 
the final form of the Hamiltonian
\beq
  H_0^{j \rightarrow \infty} = 
 \omega a^\dag a
  + \rb{
      \frac{4\lambda^2}{\omega} \pm \Omega
    } b^\dag b
  - j 
    \rb{
      \frac{2 \lambda^2}{\omega} \pm \Omega
    }.
\eeq
The excitation energies of this Hamiltonian are clearly always real.  
However, only the Hamiltonian with the upper sign 
(corresponding to $\sqrt{\beta} = + \sqrt{j}$) has the same excitation 
and ground-state energies as our previous analytic calculation.  
The solution with 
$\sqrt{\beta} = -\sqrt{j}$ leads to a Hamiltonian with
the incorrect energies, and is thus seen to be spurious.  
This solution is obviously unphysical
for $\lambda < \sqrt{\omega \Omega} /2$, as here the coefficient of the 
second oscillator becomes negative.
The origin of this spurious solution can 
be easily understood by considering the $\theta = \pi$ limit of 
the Hamiltonian.  In this case, the Hamiltonian is the same as
that of Eq. (\ref{H0}), except that $\Omega$ is replaced by $-\Omega$.
Exchanging $J_x$ for its eigenvalue as above and diagonalising the 
atomic mode, we obtain the energies
\begin{eqnarray}
  E_{n,m_x}= \omega n + -\Omega m_x - 2 \frac{\lambda^2 m_x^2}{j \omega}.
\end{eqnarray}
The problem with this Hamiltonian arises when we take the thermodynamic 
limit under the assumption that $m_x=-j$ is the spin- quantum number 
of the ground state.  This leads to the energy
\begin{eqnarray}
  E_{n,k_x}^{j\rightarrow \infty} =
  \omega n 
  + \rb{-\Omega + 4 \lambda^2 / \omega} k_x
  - j \rb{-\Omega + 2\lambda^2/\omega},
\end{eqnarray}
which is the same as the spurious solutions obtained above.  Clearly, 
the correct ground-state of the $\theta = 0$ Hamiltonian 
actually has the quantum number $m_x=+j$.  So we see that the origin of 
this type of spurious solution is due to the incorrect counting of the 
states labelled with $m_x$ as we go to the thermodynamic limit.
The solutions with the incorrect sign always have a ground-state 
energy that is higher than the correct solution, and thus we are 
easily able to discard the solutions which arise from misidentifying the 
ground state.

\begin{figure}[th]
  \centerline{
    \includegraphics[clip=true,width=1\columnwidth]{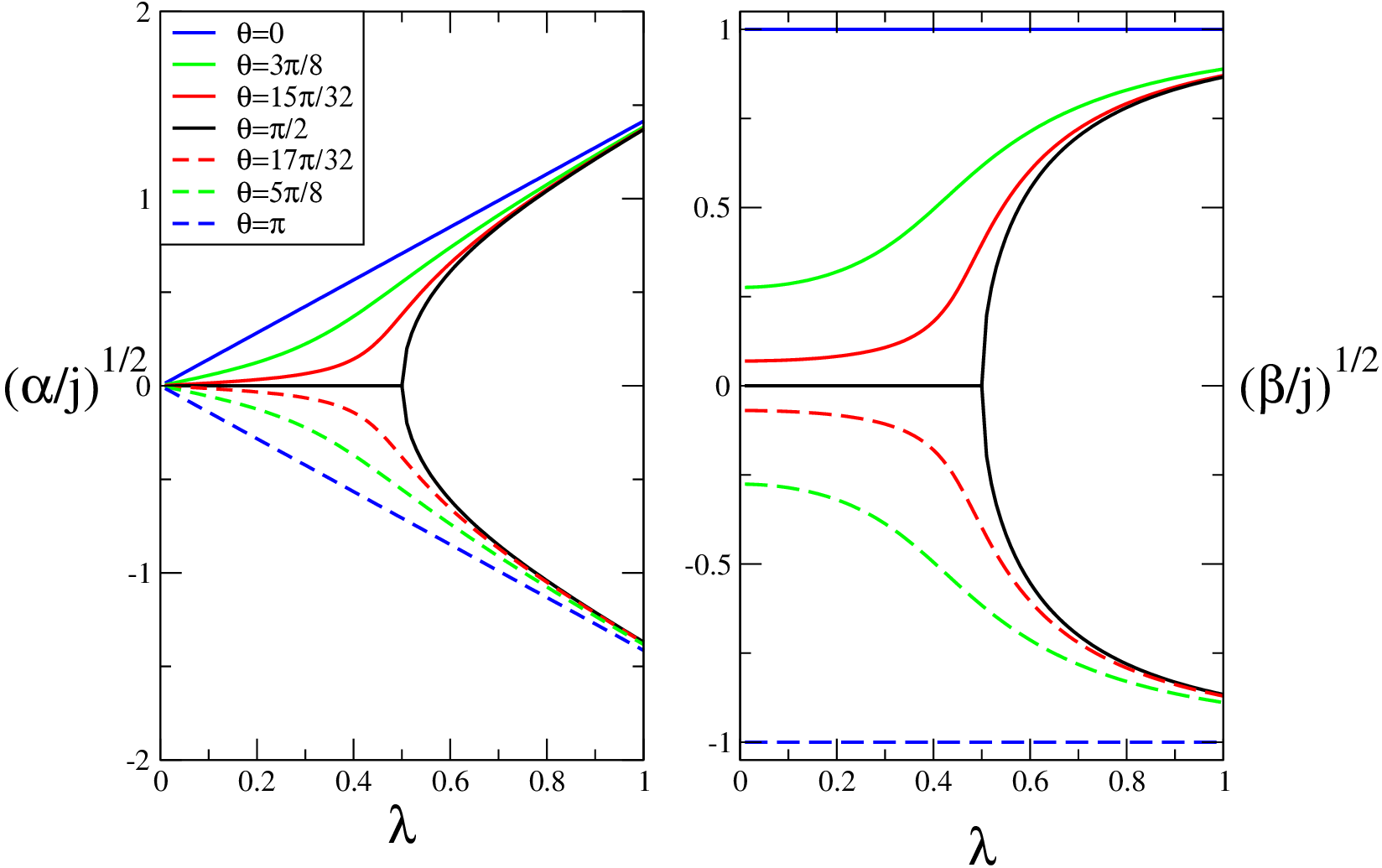}
  }
  \caption{The two displacement parameters $\sqrt{\alpha}$ 
           and $\sqrt{\beta}$ 
           as a function
           of the coupling $\lambda$ for various different angles $\theta$.
           The Hamiltonian is on scaled resonance, 
           $\omega = \Omega=1$, $\lambda_c=0.5$
           \label{disp1}
          }
\end{figure}
\begin{figure}[th]
  \centerline{
    \includegraphics[clip=true,width=1\columnwidth]{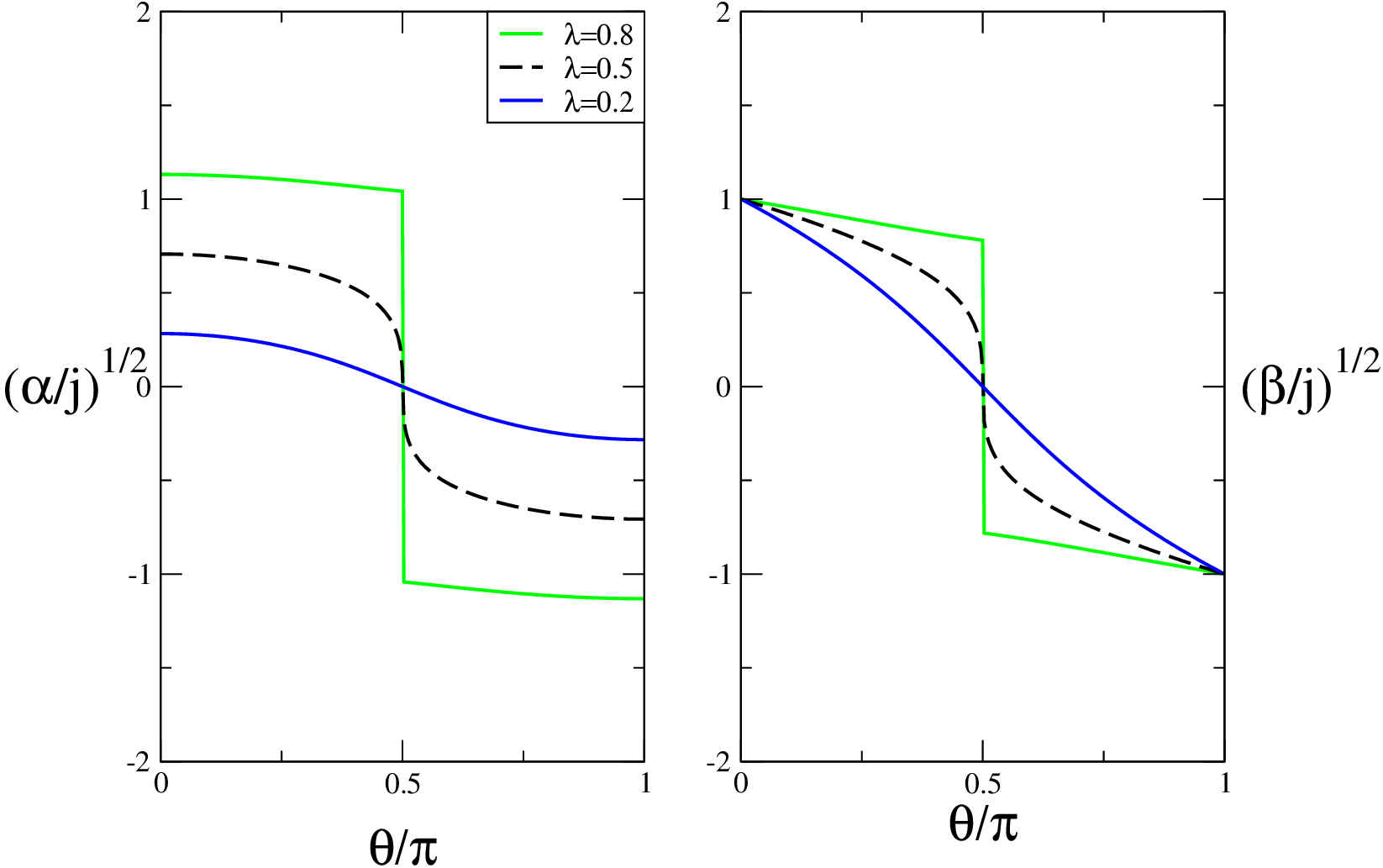}
  }
  \caption{The two displacement parameters $\sqrt{\alpha}$ and $\sqrt{\beta}$
           a function
           of the $\theta$ for representative couplings.
           The Hamiltonian is on scaled resonance,
           $\omega = \Omega=1$, $\lambda_c=0.5$
           \label{disp2}
          }
\end{figure}

\section{Results and Discussion}

To determine the behaviour of the system away from these two specific limits, 
we first solve for $\sqrt{\alpha}$ and $\sqrt{\beta}$.  Figures \ref{disp1} 
and \ref{disp2} show the values of these two displacement parameters 
as functions of both $\lambda$ and $\theta$.  Our first observation is
that for all $\theta \ne \pi/2$, there is only one solution for a given 
$\lambda$. Furthermore, the
sign of $\sqrt{\alpha}$ and $\sqrt{\beta}$ is given by that of 
$\cos\theta$.  
The divide between the regions of positive and negative displacements is
spanned by the special case of $\theta = \pi/2$, which is the previously 
discussed Dicke model.  In this case we have $\sqrt{\alpha}=\sqrt{\beta}=0$
below $\lambda_c$, and {\em two} solutions of opposite sign above $\lambda_c$.
The displacement parameters $\alpha$ and $\beta$ determine the centre(s) of 
the collective ground-state wave function of the coupled systems in a 
position-momentum representation of the two bosonic modes $a$ and $b$ 
\cite{EB03}.
The appearance of two solutions for  $\theta = \pi/2$ then corresponds to 
a breaking up of the wave function into two macroscopically separated parts
for $j\to \infty$. This parity breaking phase transition therefore 
occurs only at $\theta = \pi/2$ which demonstrates
that the Dicke model $H_{\pi/2}$ with its `orthogonal' coupling is unique 
within the whole class
of Hamiltonians $H_{\theta}$.
It is only in this special case that the super--radiant phase will 
exhibit macroscopically coherent (Schr\"odinger's cat) behaviour when 
$j$ remains finite.

\begin{figure}[th]
  \centerline{
    \includegraphics[clip=true,width=0.86\columnwidth]{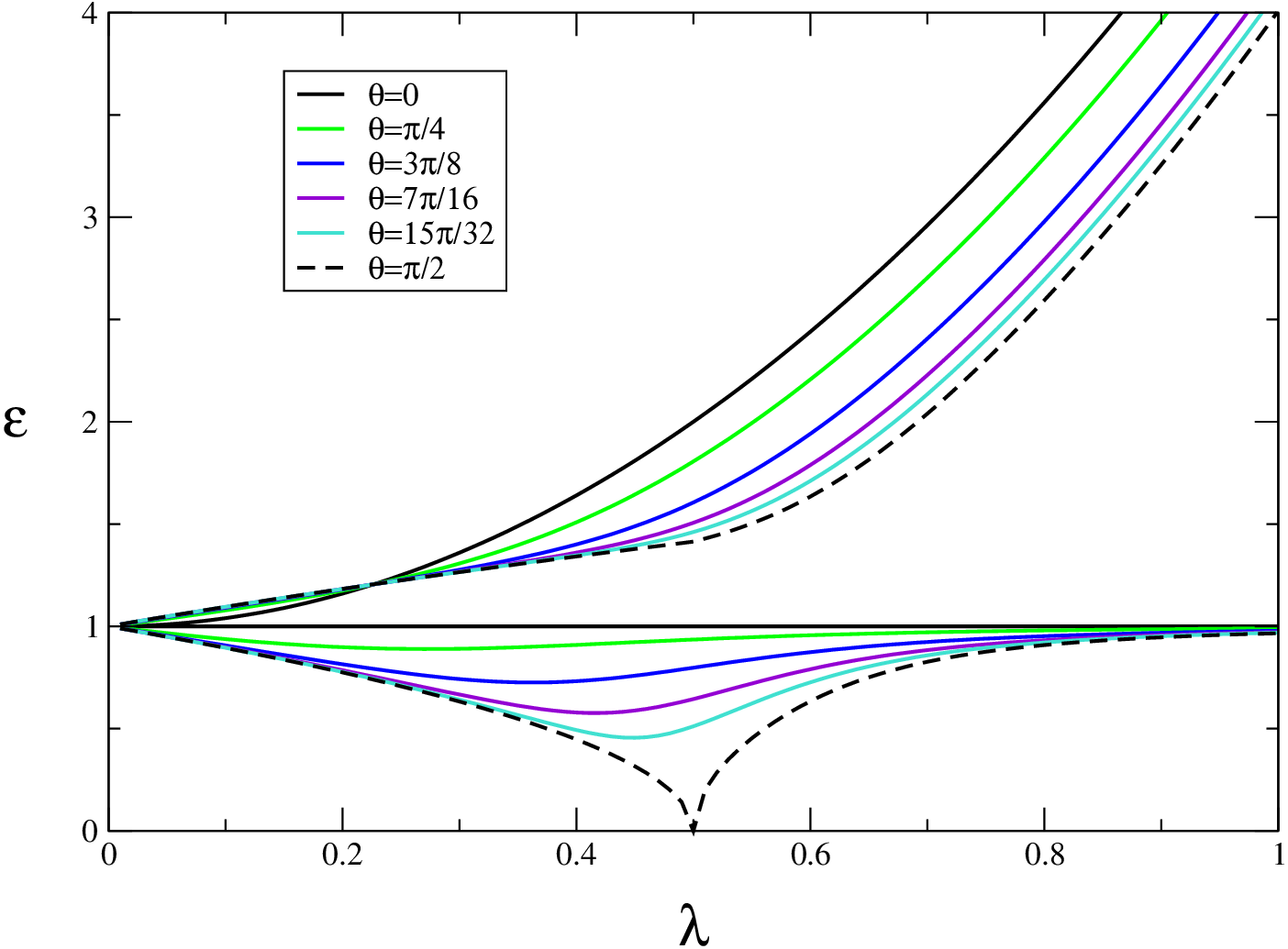}
  }
  \caption{The excitation energies of the system as a function
           of the coupling $\lambda$ for various different angles $\theta$.
           The Hamiltonian is on scaled resonance,
           $\omega = \Omega=1$, $\lambda_c=0.5$
           \label{Xenergy}
          }
\end{figure}

\begin{figure}[th]
  \centerline{
    \includegraphics[clip=true,width=0.86\columnwidth]{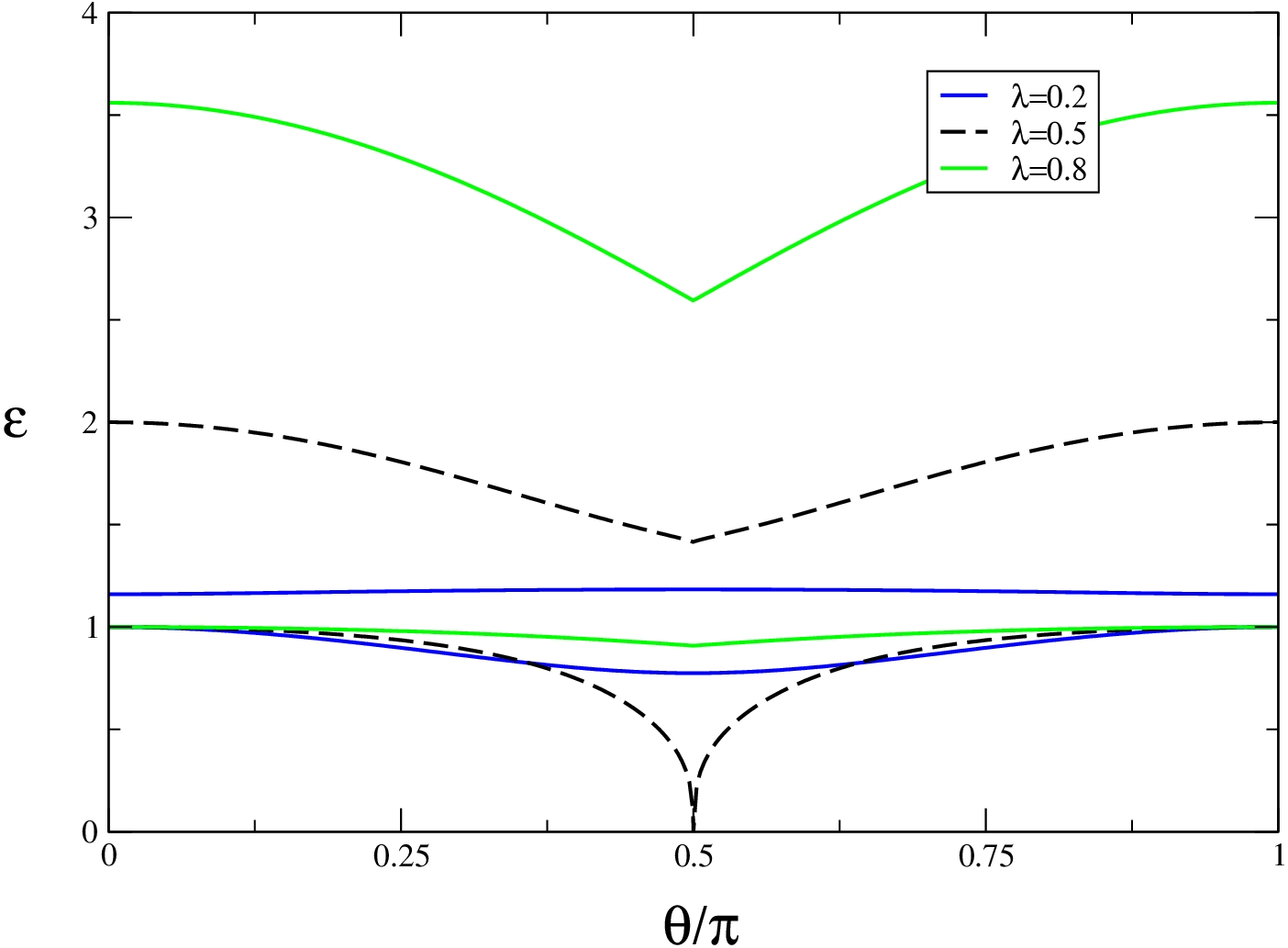}
  }
  \caption{The excitation energies of the system as a function
           of angle $\theta$ for representative values of 
           coupling $\lambda$.
           The Hamiltonian is  on scaled resonance,
           $\omega = \Omega=1$, $\lambda_c=0.5$
           \label{Xenergytheta}
          }
\end{figure}

\begin{figure}[th]
  \centerline{
    \includegraphics[clip=true,width=0.9\columnwidth]{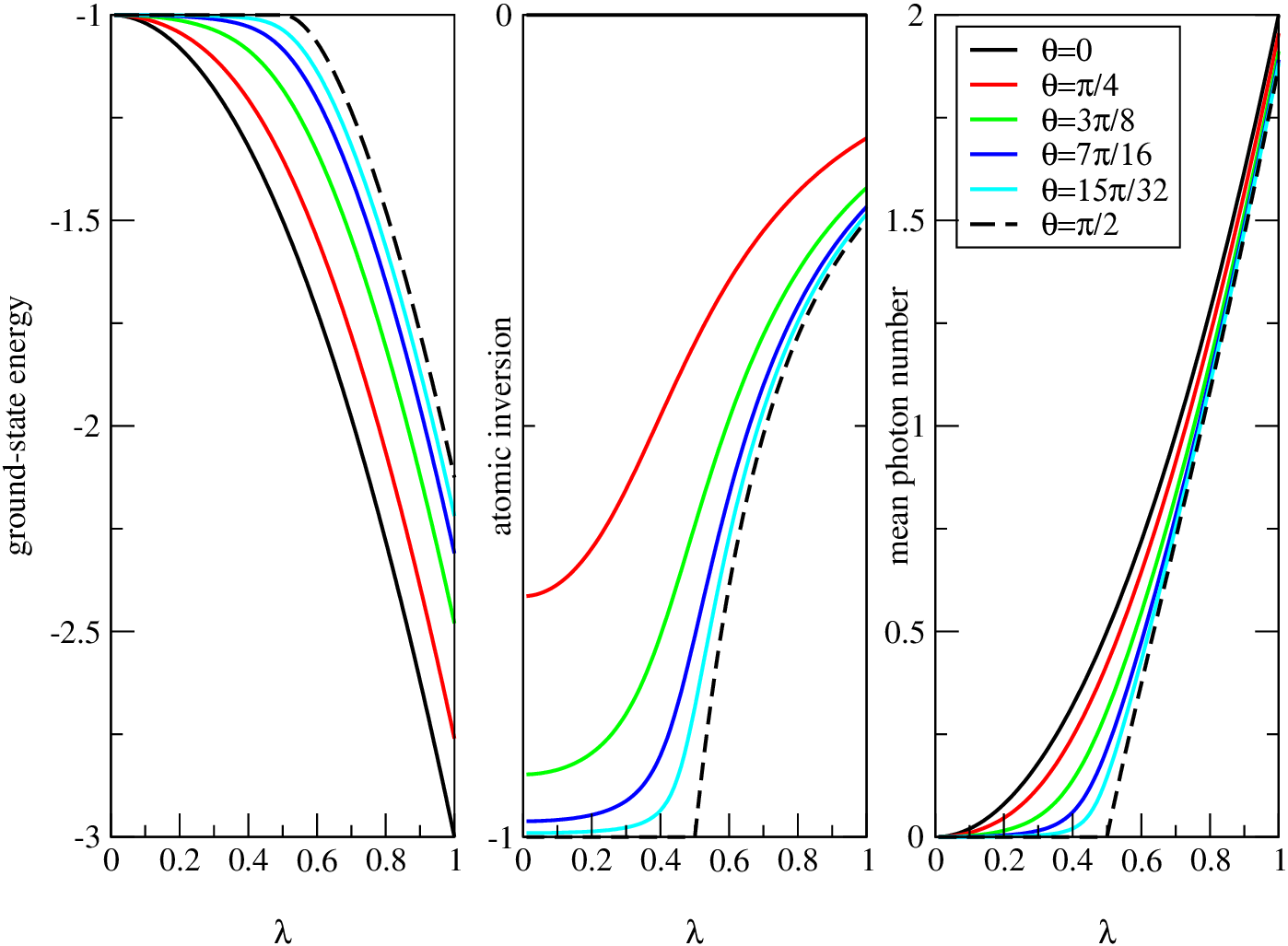}
  }
  \caption{The ground-state energy, atomic inversion and
           mean photon number of the ground state
           as a function 
           of the coupling $\lambda$ for various different angles $\theta$.
           The Hamiltonian is  on scaled resonance,
           $\omega = \Omega=1$, $\lambda_c=0.5$
           \label{observables}
          }
\end{figure}

This conclusion is corroborated by considering the excited states of our 
models.
The nature of the system is characterised by the behaviour of its
two excitation energies, which are plotted in Figs. \ref{Xenergy} and
\ref{Xenergytheta}.  In Fig. \ref{Xenergy} the limiting cases of 
$\theta =0$ and $\theta = \pi/2$ are clearly identifiable, and serve to 
provide bounds for the other solutions away from these values.  The most 
crucial consequence of this is that again, only for $\theta=\pi/2$ and 
$\lambda=\lambda_c$ does $\varepsilon_-$ identically vanish, and so it 
is only for these parameter values that a quantum phase transition occurs.

A further check is made  
in Fig.  \ref{observables}, where we plot the values of important observables of
the system.  The expression for the ground-state energy has been given 
in Eq. (\ref{gse}).  The atomic inversion and mean field occupation 
are given by
\beq
  \ew{J_z}/j = \beta/j -1
  ;~~~
  \ew{a^\dag a} = \alpha/j.
\eeq
Again, singular behaviour in the form of non-analyticities 
of the curves at $\lambda=\lambda_c=1/2$
is observed only at $\theta=\pi/2$ in agreement with the 
above result.

To summarise, the existence of the 
quantum phase transition for  spin-boson models $H_{\theta}$  
 is dependent on the two vectors $\Lambda$
and $\Omega$ being exactly perpendicular, which one might not have 
expected at the outset.
In conclusion, we briefly discuss the implications  these findings have for 
spin-boson systems. One obvious consequence is that
`non-orthogonal' coupling terms always would smear out  phase transitions 
or their precursors when tuning from a  
weak to a strong coupling regime in, e.g., photon or phonon cavities. At 
first sight, this
looks like bad news for the possible realisation of critical behavior in 
realistic 
systems where one would always expect perturbative terms leading to  
a general, not necessarily orthogonal coupling, unless some symmetry prevents 
this from occuring. 
On the other hand, it would be desirable to explore tunable systems where one
can vary the parameter $\theta$ (for example by using
external electric or magnetic fields),  in order to 
test some of our predictions.

\begin{acknowledgments}
This work was supported by projects EPSRC GR44690/01, DFG Br1528/4-1,
the WE Heraeus foundation, and the Dutch Science Foundation NWO/FOM.
\end{acknowledgments}


\begin{thebibliography}{20}
\expandafter\ifx\csname natexlab\endcsname\relax\def\natexlab#1{#1}\fi
\expandafter\ifx\csname bibnamefont\endcsname\relax
  \def\bibnamefont#1{#1}\fi
\expandafter\ifx\csname bibfnamefont\endcsname\relax
  \def\bibfnamefont#1{#1}\fi
\expandafter\ifx\csname citenamefont\endcsname\relax
  \def\citenamefont#1{#1}\fi
\expandafter\ifx\csname url\endcsname\relax
  \def\url#1{\texttt{#1}}\fi
\expandafter\ifx\csname urlprefix\endcsname\relax\def\urlprefix{URL }\fi
\providecommand{\bibinfo}[2]{#2}
\providecommand{\eprint}[2][]{\url{#2}}

\bibitem[{\citenamefont{Walls and Milburn}(1994)}]{Walls}
\bibinfo{author}{\bibfnamefont{D.~F.} \bibnamefont{Walls}} \bibnamefont{and}
  \bibinfo{author}{\bibfnamefont{G.~J.} \bibnamefont{Milburn}},
  \emph{\bibinfo{title}{Quantum Optics}} (\bibinfo{publisher}{Springer},
  \bibinfo{address}{Berlin}, \bibinfo{year}{1994}).

\bibitem[{\citenamefont{{A. Klein and E. R. Marshalek}}(1991)}]{KM91}
\bibinfo{author}{\bibnamefont{{A. Klein and E. R. Marshalek}}},
  \bibinfo{journal}{Rev.~Mod.~Phys.} \textbf{\bibinfo{volume}{63}},
  \bibinfo{pages}{375} (\bibinfo{year}{1991}).

\bibitem[{\citenamefont{Haake}(2001)}]{Haake}
\bibinfo{author}{\bibfnamefont{F.}~\bibnamefont{Haake}},
  \emph{\bibinfo{title}{Quantum Signatures of Chaos}}
  (\bibinfo{publisher}{Springer}, \bibinfo{address}{Berlin, Heidelberg},
  \bibinfo{year}{2001}).

\bibitem[{\citenamefont{Weiss}(1993)}]{Weiss}
\bibinfo{author}{\bibfnamefont{U.}~\bibnamefont{Weiss}},
  \emph{\bibinfo{title}{Quantum Dissipative Systems}}, vol.~\bibinfo{volume}{2}
  of \emph{\bibinfo{series}{Series of Modern Condensed Matter Physics}}
  (\bibinfo{publisher}{World Scientific}, \bibinfo{address}{Singapore},
  \bibinfo{year}{1993}).

\bibitem[{\citenamefont{{H. Park, J. Park, A. K. L. Lim, E. H. Anderson, A. P.
  Alivisatos, and P. L. McEuen}}(2000)}]{Paretal00}
\bibinfo{author}{\bibnamefont{{H. Park, J. Park, A. K. L. Lim, E. H. Anderson,
  A. P. Alivisatos, and P. L. McEuen}}}, \bibinfo{journal}{Nature}
  \textbf{\bibinfo{volume}{407}}, \bibinfo{pages}{57} (\bibinfo{year}{2000}).

\bibitem[{\citenamefont{{A.~N.~Cleland, M.~L.~Roukes}}(1998)}]{CR98}
\bibinfo{author}{\bibnamefont{{A.~N.~Cleland, M.~L.~Roukes}}},
  \bibinfo{journal}{Nature} \textbf{\bibinfo{volume}{392}},
  \bibinfo{pages}{160} (\bibinfo{year}{1998}).

\bibitem[{\citenamefont{{R. H. Blick, F. G. Monzon, W. Wegscheider, M. Bichler,
  F. Stern, and M. L. Roukes}}(2000)}]{Blietal00}
\bibinfo{author}{\bibnamefont{{R. H. Blick, F. G. Monzon, W. Wegscheider, M.
  Bichler, F. Stern, and M. L. Roukes}}}, \bibinfo{journal}{Phys. Rev. B}
  \textbf{\bibinfo{volume}{62}}, \bibinfo{pages}{17103} (\bibinfo{year}{2000}).

\bibitem[{\citenamefont{{T. Brandes, N. Lambert}}(2003)}]{BL03}
\bibinfo{author}{\bibnamefont{{T. Brandes, N. Lambert}}},
  \bibinfo{journal}{Phys. Rev. B} \textbf{\bibinfo{volume}{67}},
  \bibinfo{pages}{125323} (\bibinfo{year}{2003}).

\bibitem[{\citenamefont{{T. Holstein and H. Primakoff}}(1949)}]{HP49}
\bibinfo{author}{\bibnamefont{{T. Holstein and H. Primakoff}}},
  \bibinfo{journal}{Phys. Rev.} \textbf{\bibinfo{volume}{58}},
  \bibinfo{pages}{1098} (\bibinfo{year}{1949}).

\bibitem[{\citenamefont{{C. Emary, T. Brandes}}(2003{\natexlab{a}})}]{EB03}
\bibinfo{author}{\bibnamefont{{C. Emary, T. Brandes}}}, \bibinfo{journal}{Phys.
  Rev. Lett.} \textbf{\bibinfo{volume}{90}}, \bibinfo{pages}{044101}
  (\bibinfo{year}{2003}{\natexlab{a}}).

\bibitem[{\citenamefont{{C. Emary, T. Brandes}}(2003{\natexlab{b}})}]{EB03a}
\bibinfo{author}{\bibnamefont{{C. Emary, T. Brandes}}}, \bibinfo{journal}{Phys.
  Rev. E} \textbf{\bibinfo{volume}{67}}, \bibinfo{pages}{066203}
  (\bibinfo{year}{2003}{\natexlab{b}}).

\bibitem[{\citenamefont{{N. Lambert, C. Emary, T. Brandes}}(2003)}]{LEB03}
\bibinfo{author}{\bibnamefont{{N. Lambert, C. Emary, T. Brandes}}},
  \bibinfo{journal}{cond-mat/0309027 (to appear in Phys. Rev. Lett.)}
  \textbf{\bibinfo{volume}{xx}}, \bibinfo{pages}{xxx} (\bibinfo{year}{2003}).

\bibitem[{\citenamefont{Dicke}(1954)}]{Dic54}
\bibinfo{author}{\bibfnamefont{R.~H.} \bibnamefont{Dicke}},
  \bibinfo{journal}{Phys. Rev.} \textbf{\bibinfo{volume}{93}},
  \bibinfo{pages}{99} (\bibinfo{year}{1954}).

\bibitem[{\citenamefont{Allen and Eberly}(1987)}]{Allen}
\bibinfo{author}{\bibfnamefont{L.}~\bibnamefont{Allen}} \bibnamefont{and}
  \bibinfo{author}{\bibfnamefont{J.~H.} \bibnamefont{Eberly}},
  \emph{\bibinfo{title}{Optical Resonance and Two-Level Atoms}}
  (\bibinfo{publisher}{Dover}, \bibinfo{address}{New York},
  \bibinfo{year}{1987}).

\bibitem[{\citenamefont{{G. M. Palma, K.-A. Suominen, A. K.
  Ekert}}(1996)}]{PSE96}
\bibinfo{author}{\bibnamefont{{G. M. Palma, K.-A. Suominen, A. K. Ekert}}},
  \bibinfo{journal}{Proc. Roy. Soc. Lond. A} \textbf{\bibinfo{volume}{452}},
  \bibinfo{pages}{567} (\bibinfo{year}{1996}).

\bibitem[{\citenamefont{{J. H. Reina, L. Quiroga, and N. F.
  Johnson}}(2002)}]{RQJ02}
\bibinfo{author}{\bibnamefont{{J. H. Reina, L. Quiroga, and N. F. Johnson}}},
  \bibinfo{journal}{Phys. Rev. A} \textbf{\bibinfo{volume}{65}},
  \bibinfo{pages}{032326} (\bibinfo{year}{2002}).

\bibitem[{\citenamefont{{T. Yu and J. H. Eberly}}(2003)}]{YE03}
\bibinfo{author}{\bibnamefont{{T. Yu and J. H. Eberly}}},
  \bibinfo{journal}{Phys. Rev. B} \textbf{\bibinfo{volume}{68}},
  \bibinfo{pages}{165322} (\bibinfo{year}{2003}).

\bibitem[{\citenamefont{{A. J. Leggett, S. Chakravarty, A. T. Dorsey, M. P. A.
  Fisher, A. Garg, and W. Zwerger}}(1987)}]{Legetal87}
\bibinfo{author}{\bibnamefont{{A. J. Leggett, S. Chakravarty, A. T. Dorsey, M.
  P. A. Fisher, A. Garg, and W. Zwerger}}}, \bibinfo{journal}{Review of Modern
  Physics} \textbf{\bibinfo{volume}{59}}, \bibinfo{pages}{1}
  (\bibinfo{year}{1987}).

\bibitem[{\citenamefont{{T. Brandes and B. Kramer, Phys. Rev. Lett. {\bf 83},
  3021 (1999); Physica B {\bf 272}, 42 (1999); Physica B {\bf 284-288},
  1774}}(2000)}]{Brandesddot}
\bibinfo{author}{\bibnamefont{{T. Brandes and B. Kramer, Phys. Rev. Lett. {\bf
  83}, 3021 (1999); Physica B {\bf 272}, 42 (1999); Physica B {\bf 284-288},
  1774}}} (\bibinfo{year}{2000}).

\bibitem[{\citenamefont{{S. Debald, private communication}}(2003)}]{DB03}
\bibinfo{author}{\bibnamefont{{S. Debald, private communication}}}
  (\bibinfo{year}{2003}).

\end{thebibliography}

\end{document}